\documentclass[prl,twocolumn,showpacs,amsmath,superscriptaddress,psfig,amssymb]{revtex4}
\usepackage{amssymb}

\usepackage{graphicx}
\usepackage{dcolumn}
\usepackage{bm}

\begin{document}


\title{Disruption of the accidental Dirac semimetal state in ZrTe$_{5}$ under hydrostatic pressure}

\author{J. L. Zhang}
\affiliation{High Magnetic Field Laboratory, Chinese Academy of Sciences, Hefei 230031, Anhui, People¡¯s Republic of China}
\author{C. Y. Guo }
\affiliation{Department of Physics and Center for Correlated Matter, Zhejiang University, Hangzhou,
Zhejiang 310027, China}
\author{X. D. Zhu}
\affiliation{High Magnetic Field Laboratory, Chinese Academy of Sciences, Hefei 230031, Anhui, People¡¯s Republic of China}
\author{L. Ma }
\affiliation{High Magnetic Field Laboratory, Chinese Academy of Sciences, Hefei 230031, Anhui, People¡¯s Republic of China}
\author{G. L. Zheng }
\affiliation{High Magnetic Field Laboratory, Chinese Academy of Sciences, Hefei 230031, Anhui, People¡¯s Republic of China}
\author{Y. Q. Wang }
\affiliation{High Magnetic Field Laboratory, Chinese Academy of Sciences, Hefei 230031, Anhui, People¡¯s Republic of China}
\author{L. Pi}
\affiliation{High Magnetic Field Laboratory, Chinese Academy of Sciences, Hefei 230031, Anhui, People¡¯s Republic of China}
\author{Y. Chen }
\affiliation{Department of Physics and Center for Correlated Matter, Zhejiang University, Hangzhou, Zhejiang 310027, China}
\author{H. Q. Yuan }
\affiliation{Department of Physics and Center for Correlated Matter, Zhejiang University, Hangzhou, Zhejiang 310027, China}
\affiliation{Collaborative Innovation Center of Advanced Microstructures, Nanjing University, Nanjing 210093, People¡¯s Republic of China}
\author{M. L. Tian}
\email{tianml@hmfl.ac.cn}
\affiliation{High Magnetic Field Laboratory, Chinese Academy of Sciences, Hefei 230031, Anhui, People¡¯s Republic of China}
\affiliation{Collaborative Innovation Center of Advanced Microstructures, Nanjing University, Nanjing 210093, People¡¯s Republic of China}

\date{\today}

\begin{abstract}

We study the effect of hydrostatic pressure on the magnetotransport properties of the zirconium pentatelluride. The magnitude of resistivity anomaly gets enhanced with increasing pressure, but the transition temperature $T^{\ast}$ is almost independent of it. In the case of H $\parallel$ $b$, the quasi-linear magnetoresistance decreases drastically from 3300$\%$ (9 T) at ambient pressure to 400$\%$ (14 T) at 2.5 GPa. Besides, the change of the quantum oscillation phase from topological nontrivial to trivial is revealed around 2 GPa.
Both demonstrate that the pressure breaks the accidental Dirac node in ZrTe$_{5}$. For H $\parallel$ $c$, in contrast, subtle changes can be seen in the magnetoresistance and quantum oscillations. In the presence of pressure, ZrTe$_{5}$ evolves from a highly anisotropic to a nearly isotropic electronic system, which accompanies with the disruption of the accidental Dirac semimetal state. It supports the assumption that ZrTe$_{5}$ is a semi-3D Dirac system with linear dispersion along two directions and a quadratic one along the third.

\end{abstract}


\maketitle

In past few years, the topological quantum materials such as topological insulators (TIs) \cite{QiXL,Hasan},
Dirac semimetals \cite{WangZJ,WangZJ2013,LiuZK,LiuZKSci} and Weyl semimetals \cite{Weng2015,XuSY,LvBQ,HuangXC,ZhangCL} have stimulated unprecedented research interest both theoretically and experimentally for
their unique electronic states. Layered compound ZrTe$_{5}$ has been studied for decades
due to its large thermoelectric power, mysterious resistivity anomaly and large positive magnetoresistance \cite{Jones, Rubinstein, Tritt}.
In 2014, Weng $et$ $al$. predicted that the single-layer ZrTe$_{5}$ is a candidate of large-gap quantum spin Hall insulator \cite{Weng}. On the other hand, the 3D bulk crystal is either a weak or a strong topological insulator,
in which the interlayer spacing is believed to play a key role in defining its topological character \cite{Weng}.
Experimental verification, however, remains highly controversial. Scanning tunneling microscopy/
spectroscopy (STM/STS) and angle-resolved photoemission spectroscopy (ARPES) measurements detected a bulk band gap with topological edge states at the surface step edge \cite{WuR,LiXB}, hosting the signatures of a weak 3D TI. In contrast,
the chiral magnetic effect and non-trivial Berry phase was clearly observed in ZrTe$_{5}$
through magneto-transport measurements \cite{LiQ,ZhengGL,YuanX,LiuYW}, and the ARPES experiments further identified it to be a
3D Dirac semimetal with only one Dirac node at the $\Gamma$ point \cite{LiQ, ShenL}. In addition, magnetoinfrared
spectroscopy results, such as the linear energy dependence of optical conductivity
and Landau level splitting, also support this scenario \cite{ChenRYPRB,ChenRYPRL}. Very recently, Manzoni $et$ $al$. find out that the 3D Dirac semimetal phase manifests  at the boundary between the weak and strong TI phases \cite{Manzoni}.

Applying pressure is known to be a powerful approach to tune the electronic states and lattice structures without introducing disorder or impurity,
which has been widely employed in topological materials. Recently, a pressure-induced semimetal to superconductor transition was observed
in ZrTe$_{5}$ at 6.2 GPa \cite{ZhouYH}. However, no quantum oscillation is observed in their pressure study, which is likely due to the pressure inhomogeneity. Since quantum oscillations can provide direct evidence of the topological nature of a material, further experiment is desired on this approach.

In this letter, we study the magnetoresistance and S-dH quantum oscillations
for ZrTe$_{5}$ single crystals under hydrostatic pressure. For H $\parallel$ $b$, the quasi-linear large magnetoresistance is suppressed with pressure.
Meanwhile, an abrupt quantum oscillation phase shift is observed around 2 GPa, indicating there is a pressure-induced topological phase transition.
These results suggest the disruption of the Dirac semimetal state under pressure.
In the case of H $\parallel$ $c$, both the magnetoresistance and topological features
of Fermi surface in the $ab$ plane change slightly with pressure.
Moreover, the anisotropic parameter of quantum oscillation frequency $F^{ab}/F^{ac}$ (effective mass $m^{ab}/m^{ac}$)
decreases monotonically with pressure and approaches to 1.5 (1) at 2.5 GPa, implying that electronic structure of ZrTe$_{5}$ tends to be isotropic.

\begin{figure}[tb]\centering
 \includegraphics[width=9cm]{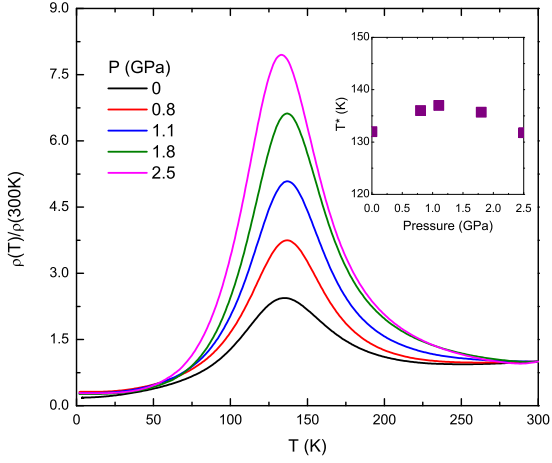}
\caption{Temperature dependence of
the electrical resistivity $\rho(T)$ for ZrTe$_{5}$ in zero magnetic field under selected pressures. Data are normalized to the value of resistivity at 300 K.
The inset shows the evolution of resistivity anomaly $T^{*}$ under pressure.}\label{fig.1}
\end{figure}

\begin{figure}[tb]\centering
 \includegraphics[width=9cm,height=10cm]{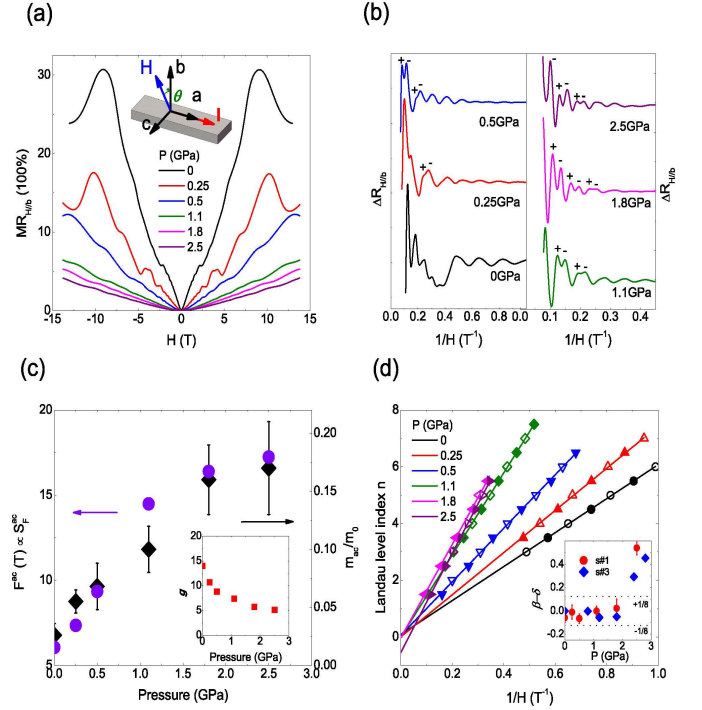}
\caption{(a) the magnetoresistance of ZrTe$_{5}$ at various pressures with H $\parallel b$ and $\mathbf{I} \parallel a$, measured at 0.3 K. Inset schematically displays the arrangements of magnetic field and current applied for sample$\#$1. (b) Shubnikov-de Haas oscillations component $\Delta$R$_{H\parallel c}$ as a function of inverse magnetic field taken at different pressures. Data are normalized and shifted for clarity. (c) Pressure dependence of the quantum oscillation frequency (Fermi surface size) and the effective mass. Inset: pressure dependence of the Landau $g-$factor. (d) Landau level index plots of the oscillations at various pressure. Inset summarized the phase factor of S-dH oscillation as a function of pressure. Circle and diamond represents for sample$\#$1 and sample$\#$3 respectively. }\label{fig.3}
\end{figure}

High quality single crystals of ZrTe$_{5}$ used in our studies were
 synthesized using the iodine vapor transport method in a two-zone furnace.
 Details of the crystal-growth procedure are given in Ref \cite{Kam}.
 Electrical-transport experiments were performed in a Physical
 Property Measurement System (PPMS quantum design inc) from 2 K to 300 K
 and a $^{3}$He refrigerator down to 0.3 K. Resistivity measurements were
 carried by the standard four-probe method, where the contacts were
 made with spot welding and each contact resistance is less than
 5 $\Omega$. Two samples were cut from the same piece of a single
 crystal. Typical dimensions of the needle-like samples were about
 1 mm$\times$0.1 mm$\times$0.2 mm. The magnetic field was parallel to
 the $b$-axis for sample$\#1$ and $c$-axis for sample$\#2$,
 respectively. In both cases, electrical current $\mathbf{I}$ was applied along
  the $a$-axis, perpendicular to the magnetic field. Hydrostatic pressure
  was generated using a piston-clamp type pressure cell utilizing with
  Daphne 7373 oil as pressure transmitting medium. The pressure inside
  the cell was determined by measuring the superconducting transition
  temperature of high quality Pb.

Fig. 1 represents temperature dependence of the normalized resistivity for ZrTe$_{5}$
single crystal at various pressures. At ambient pressure, the resistivity
shows a broad peak at $T^{\ast}$=132 K, which is known as the hallmark of ZrTe$_{5}$ \cite{Rubinstein,ZhengGL,YuanX}. Recent ARPES studies found out that the resistivity peak
results from a temperature-induced Lifshitz-transition where ZrTe$_{5}$ evolves
from semiconductor to n-type semimetal \cite{ZhangY}. The magnitude of resistivity anomaly gets enhanced with increasing pressure.
It indicates that the pressure enlarges the direct gap between the valance and conduction bands, which is consistent with band structure calculations \cite{FanZJ}.
On the other hand, as shown in the inset of Fig. 1, $T^{\ast}$
varies slightly with pressure, implying that the typical band feature of ZrTe$_{5}$ does not change drastically in this pressure range. Note that our results are in contrast
to recent high pressure study performed by diamond-anvil cell (DAC) \cite{ZhouYH}, in which the resistivity anomaly is suppressed and $T^{\ast}$ moves
 to high temperature under pressure.
Whether such discrepancy is resulted from different pressure homogeneity needs to be confirmed in further studies.

Now we turn to investigate the evolution of longitudinal magnetoresistance of ZrTe$_{5}$ under pressure.
A magnetic field was applied from $-$14 T up to 14 T. Throughout this paper,
the magnetoresistance was defined as MR = [R(H)-R(0)]/R(0)$\times$100$\%$. In order to remove the Hall contribution
from the data, MR curves are symmetrized via MR$_{symm}$(H) = 0.5$\times$[MR($-$H)+MR(H)]. Firstly, we focus on the MR$_{H \parallel b}$ where the magnetic field was applied along the $b$-axis (as illustrated in the inset of Fig. 2(a)). At ambient pressure, the MR$_{H \parallel b}$ of ZrTe$_{5}$
increases rapidly with pressure and reach to approximately 3300$\%$ at 9 T, then followed by a sudden drop whose
magnitude is much larger than that of quantum oscillation. In recent works, it was explained by the picture
of either dynamical mass generation \cite{LiuYW} or topological phase transition from 3D Weyl semimetal to 2D massive Dirac metal \cite{ZhengGL 2016}. With increasing pressure, such MR anomaly shifts to high magnetic field. Unfortunately, above 0.5 GPa, we can not directly determine it even in
the highest field we can reach. Remarkably, the magnetoresistance of ZrTe$_{5}$ is suppressed dramatically with pressure. At 2.5 GPa, the MR$_{H \parallel b}$ is around 400$\%$ in a magnetic field of 14 T, nearly one order smaller than that at ambient pressure. Similar drastic pressure effect was previously reported in type II Weyl semimetal WTe$_{2}$, in which the pressure breaks the perfect balance between the hole and electron pockets \cite{HeLP}. However, such mechanism is unlikely to explain our data, because the field dependence of MR for WTe$_{2}$ follows
quadratic behavior, while the MR of ZrTe$_{5}$ increases quite linearly with magnetic field at various pressures. On the other hand, large linear magnetoresistance had been observed in Dirac semimetals and topological insulators which are linked to the peculiar electronic states, known as Dirac fermions/dispersions \cite{QuDX,LiangT,CaiPL,Narayanan}. In this sense, the suppression of the magetoresistance could be understood from the disturbance of linear band dispersion features near the Fermi level under pressure.

\begin{figure}[tb]\centering
\includegraphics[width=9cm,height=10cm]{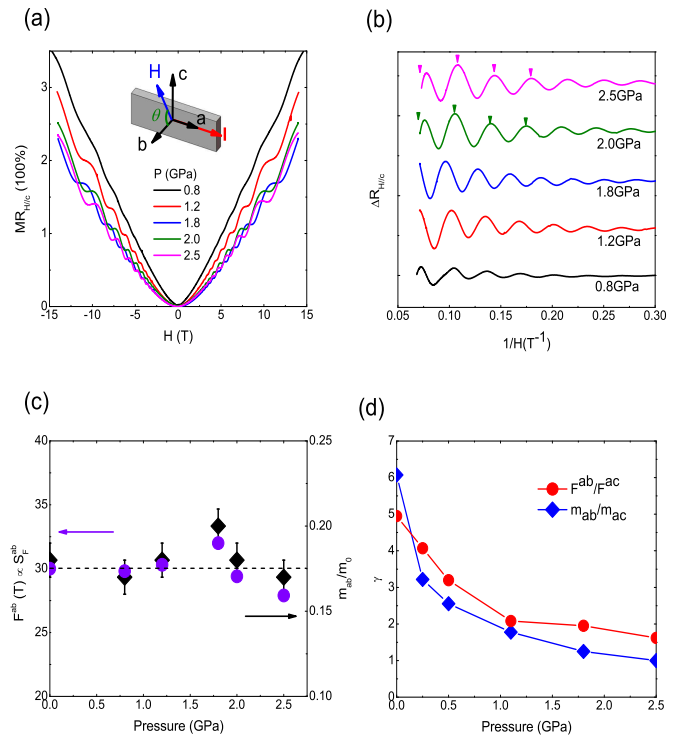}
\caption{(a) the magnetoresistance of ZrTe$_{5}$ at various pressures with H$\parallel$c and $\mathbf{I}\parallel$a, measured at 0.3 K. Inset schematically displays the arrangements of magnetic field and current applied for sample$\#$2. (b) Shubnikov-de Haas oscillations component $\Delta$R$_{H\parallel c}$ as a function of inverse magnetic field taken at different pressures. Data are shifted for clarity and arrows indicate the Landau levels. (c) Pressure dependence of the quantum oscillation frequency (Fermi surface size) and the effective mass. (d) Evolution of the anisotropic parameters under pressure. }\label{fig.3}
\end{figure}

Measurement of quantum oscillations is a very powerful way of determining the topological order of the Fermi surface (FS). The MR$_{H \parallel b}$ of ZrTe$_{5}$ exhibits pronounced
S-dH quantum oscillations, whose amplitudes decrease gradually with increasing pressure. After subtracting the
background by using polynomials fits, oscillation patterns of ZrTe$_{5}$ at selected pressures are presented in
Fig. 2(b). Curves are equally scaled and shifted for clarity. Based on the Lifshitz$-$Kosevich (LK) formula \cite{Murakawa,Shoenberg}, the quantum oscillation of resistance could be written approximately as:

\begin{equation}
\triangle R \propto R_{T}R_{D}R_{s}cos2\pi(\frac{F}{H}-\gamma+\delta),
\end{equation}

where $F$ is the quantum oscillation frequency, $R_{T}$, $R_{D}$ and $R_{s}$ are three reduction factors accounting for temperature, scattering
and spin splitting, respectively. $\gamma-\delta$ is a phase factor which will be
discussed later. At ambient pressure, owing to the high mobility of our sample, the oscillations start
at very low magnetic field H = 0.5 T. Applying a fast Fourier transform (FFT) we obtain only a single frequency $F$ = 6 T, slightly
higher than the single crystalline nanoribbons \cite{ZhengGL}. According to the Onsager relation \cite{Shoenberg}, the frequency $F$ is proportional
to an extremal cross-sectional area of the FS in momentum space: $F = {\hbar}/({2\pi e)S_{F}}$, here $\hbar$ is the reduced
Planck constant and $e$ is the elementary charge. We obtain an extremely small cross section of FS, which is around 6.76$\times$10$^{-4}${\AA}$^{-2}$. Additionally, the quantum oscillation frequency varies from 6 to 20 in the pressure range from 0 to 2.5 GPa. The enhancement of quantum oscillation frequency corresponding to a increase of the FS size, indicates that the Fermi energy is shifted up with pressure. We estimate the effective mass from the temperature dependence of oscillation amplitude via $R_{T}=\alpha Tm^{*}/H$sinh$(\alpha Tm^{*}/H)$, where $m^{*}$ is the free-electron mass and $\alpha$ = 2$\pi^{2}k_Bm_e\sim$14.69 T/K.
In high pressure range, due to the strong
Zeeman effect, our estimation have relatively large error bar. Nonetheless, as shown in Fig. 2(c), $m_{ac}$ increases monotonously with pressure and its pressure dependence almost matches that of the quantum oscillation frequency.
In this field direction, the spin degeneracy is lifted with a weak magnetic field, generating a pronounced Zeeman splitting.
Landau $g$-factor could be extracted directly from the oscillation curves via $g=2F\Delta(1/H)m_{0}/m^{\ast}$ \cite{Shoenberg}, where $\Delta(1/H)$ is the spacing between the split peaks corresponding to the up and down spins on one Landau level. At p = 0 GPa, $g$-factor of ZrTe$_{5}$ could reach up to 15 in good agreement with magnetoinfrared spectroscopy results \cite{ChenRYPRL}. At high pressure, as illustrated in the inset of Fig. 2(d), the value of $g$-factor decreases.

According to the Lifshitz-Onsager
quantization rule, the
Landau index $n$ is linearly dependent on 1/H: $\hbar S_{n}/2\pi eH=n-\gamma+\delta$, here $\gamma$ is the Onsager phase factor that is related to the Berry phase through $\gamma=1/2-\phi_{B}/2\pi$ \cite{XiaoD}. In conventional electron system, Berry phase $\phi_{B}=0$, hence $\gamma=1/2$. While for those
massless Dirac materials with linear band dispersion, non-trivial Berry phase
$\phi=\pi$, which makes $\gamma=$0 \cite{Mikitik}. $\delta$ is an additional phase shift, which takes from $-1/8$ to $1/8$ depending on the
degree of the dimensionality of the FS. We define the peak positions as the Landau integer indices and the valley positions as the half indices. In order to avoid the influence
from the Zeeman effect, we only consider the valleys and the peaks without splitting. The linear extrapolation of $1/H$ versus the integer $n$ plot is presented in Fig. 2(d). Below 2 GPa, $\gamma-\delta$ locates between 0$\pm$1/8. The zero Onsager phase corresponds to a nontrivial Berry phase, suggesting the existence of a van Hove singularity close to the Fermi level. While above 2 GPa, it shifts to about 0.5, corresponding to a trivial Berry phase. Similar results are also observed in another sample (For sample$\#$3, the tilted angle between the $H$ and $b$-axis is around 30$^{\circ}$ \cite{SM}). Our results clearly demonstrate that there is a pressure-induced topological phase transition.

Considering the fact that ZrTe$_{5}$ has quasi-2D crystal structure,
we further study the pressure effect on the magnetoresistance with the magnetic field parallel to the $c$-axis. Sample$\#$2 is not perfectly cleaved in the $ac$ plane and the calibrated tilted angle $\theta$ is around 75$^{\circ}$ \cite{SM}). According to previous studies \cite{YuanX,LiuYW}, at such tilted angle, the MR and S-dH quantum oscillations can still capture the electron behavior in the $ab$ plane even if the field is not strictly alone the $c$-axis. At ambient pressure, ZrTe$_{5}$ exhibits large anisotropic magnetoresistance, i.e., the MR$_{H \parallel c}$ is only around 250$\%$ at 14 T, one order smaller than the MR$_{H \parallel b}$. Moreover, the MR$_{H \parallel c}$ is robust against hydrostatic pressure. As displayed in the Fig. 3(a), both the S-dH quantum oscillation frequency and effective mass $m_{ab}$ remain almost unaltered under pressure, demonstrating the stability of the electronic structure in the $ab$ plane. In low pressure ranges, there are no signatures of Zeeman splitting. But we need to point out that, when p $\geq$ 2 GPa, the last peak is located between two Landau levels (marked by arrows in Fig. 3(b)), which might hint that the Zeeman splitting occurs around 14 T. Further experiments in even higher magnetic field is necessary to clarify this point, which is beyond the scope of this work.

 Since it is not protected by crystalline symmetry, ZrTe$_{5}$ could not be strictly regarded as a Dirac semimetal like Cd$_{3}$As$_{2}$ and Na$_{3}$Bi \cite{Weng}. However, the direct gap between the valence and conduction bands shows a trend of closing with decreasing temperature \cite{ZhangY}, which would result in an accidental bulk Dirac cone at low temperatures \cite{Moreschini}. In the presence of the pressure, the Dirac cone is shifted away from the Fermi level and/or opens a finite energy gap. Recent works revealed that the Dirac semimetal state in ZrTe$_{5}$ is highly sensitive to the the lattice constants and manifests only at the boundary between the WTI and the STI phases \cite{Manzoni,FanZJ}. In fact, our results show that it would not be completely destroyed until the pressure up to 2.5 GPa. Additionally, we notice that evidences for 3D massless Dirac fermions could be widely observed in ZrTe$_{5}$ samples prepared with different growth methods
 or under specific growth conditions \cite{LiQ,ZhengGL,YuanX,ChenRYPRB,ChenRYPRL,ZhengGL 2016}. These imply that such accidental Dirac semimetal state is more robust
 than theoretical prediction. As displayed in Fig. 3(d), with increasing pressure, the anisotropic parameter of quantum oscillation frequency (effective mass)
decreases monotonically with pressure and approaches to 1.5 (1) at 2.5 GPa, hinting that the electronic state of ZrTe$_{5}$ evolves from highly anisotropic to nearly isotropic. Note that such evolution along with the disturbance of the Dirac semimetal state, which supporting recent assumption that ZrTe$_{5}$ is a semi-3D Dirac system with quadratic dispersion in $b-$direction and linear in the others \cite{YuanX}.

In summary, we have studied the magetoresistance and S-dH quantum oscillations of ZrTe$_{5}$ single crystal under hydrostatic pressure. At ambient pressure, the non-trivial Berry phase together with very light effective mass demonstrate the presence of
a van Hove singularity close to the Fermi level. Remarkably, with increasing pressure, the MR$_{H \parallel b}$ is drastically suppressed. Moreover, quantum oscillations show a phase shift from non-trivial Berry phase to trivial Berry phase.
These results indicate that the accidental Dirac cone is gradually broken under pressure.
On the other hand, for the field along the $c$-axis, both magnetoresistance and quantum oscillations are robust against pressure. Combining with the studies in different field orientations, we find that the electronic
structure evolved from highly anisotropic to nearly isotropic under pressure. Our experiments illustrate that hydrostatic pressure could tune the topological state in ZrTe$_{5}$ which would be helpful for the understanding of this complex layered material.

\acknowledgments

We acknowledge very helpful discussions with H. M. Weng, Z. J. Xiang, C. Zhang and L. Jiao. This work was
supported by the Natural Science Foundation of China (Grants No. 11204312, No. 11374302, No. 11474289,
No. U1432251, No. U1632275, No. 11504377, No. 11504378); Natural Science Foundation of Anhui Province (No. 1608085QA16); and the program
of Users with Excellence, the Hefei Science Center of CAS.

\end{document}